\documentclass[ aps, showpacs, showkeys, nofootinbib, floatfix, superscriptaddress]{revtex4}

\usepackage{amsfonts}

\usepackage{amssymb}
\usepackage{amsmath}
\usepackage{graphicx}

\begin{document}

\title{Investigate the interaction between dark matter and dark energy}

 \author{Jianbo Lu}
 \email{lvjianbo819@163.com}
 \affiliation{Department of Physics, Liaoning Normal University, Dalian 116029, P. R. China}
 \author{Yabo Wu}
 \affiliation{Department of Physics, Liaoning Normal University, Dalian 116029, P. R. China}
 \author{Yongyi Jin}
 \affiliation{China Criminal Police University, Shenyang, P. R. China}
 \author{Yan Wang}
 \affiliation{Department of Physics, Liaoning Normal University, Dalian 116029, P. R. China}

\begin{abstract}
 In this paper we investigate the interaction between dark matter and dark energy by considering
 two different interacting scenarios, i.e.  the cases of
  constant interaction function and  variable interaction
 function.
 By fitting the current observational
 data to constrain the interacting models, it is found that the interacting strength is non-vanishing,
 but weak for the
 case of constant interaction function, and the interaction is not obvious for the case of variable
 interaction function. In addition, for seeing the influence from interaction
 we also investigate the evolutions
 of interaction function, effective state parameter for dark energy and
 energy density of dark matter. At last some geometrical quantities in the interacting scenarios are discussed.
\end{abstract}



\maketitle

\section{$\text{Introduction}$}

{The observation of the supernovae of type Ia \cite{SNIa,SNIa1}
provides the evidence that the universe is undergoing accelerated
expansion. In theory, a popular interpretation for this phenomenon
is that an unknown  fluid with negative pressure, dubbed dark
energy, is introduced in universe in the framework of standard
cosmology. Many dark energy models
\cite{DEmodels0,DEmodels1,DEmodels2,DEmodels3,DEmodels4,DEmodels5,DEmodels6,
DEmodels7,DEmodels8,DEmodels9,DEmodels10,DEmodels11,DEmodels12,DEmodels13,DEmodels14,DEmodels15,
DEmodels16,DEmodels17,DEmodels18,DEmodels19} have been investigated
in the recent years from  different points of view such as
cosmological constant, the fields of theory, holographic theory and
string theory etc. Though the cosmological constant model is
consistent with the current astronomy observations at $2\sigma$
confidence level, it suffers from the fine tuning and the
coincidence problems. One of the solutions to alleviating the above
two problems is to consider the interaction between the two dark
sectors of dark energy and dark matter.  Several forms of the
interacting parameter $\Gamma$ have been studied
\cite{interaction-proof1,interaction-proof2,interaction-proof3,interaction-proof4,interaction-proof5,interaction1,
interaction2,interaction3,interaction4,interaction5,interaction6},
such as $\Gamma=-\lambda \dot{\rho}_{de}$, $\Gamma=-\lambda
H\rho_{de}$, $\Gamma=-\lambda H(\rho_{dm}+\rho_{de})$
 etc. In this paper, using the
current observational data we investigate the interaction between
dark sections with a different method introduced in Ref. \cite{Fr}.

\section{$\text{Interacting dark model}$}

Considering three equations of conservation for baryon, dark matter,
and dark energy, respectively
\begin{equation}
\dot{\rho}_{b}+3H(\rho_{b}+p_{b})=0,\label{C-baryon}
\end{equation}
\begin{equation}
\dot{\rho}_{dm}+3H\gamma_{dm}^{e}\rho_{dm}=0,\label{C-dm}
\end{equation}
\begin{equation}
\dot{\rho}_{de}+3H\gamma_{de}^{e}\rho_{de}=0,\label{C-de}
\end{equation}
with the introduced effective barotropic indexes $\gamma_{i}^{e}$
\cite{Fr},
\begin{equation}
\gamma_{dm}^{e}=\gamma_{dm}+\frac{\gamma_{de}}{r}+\frac{\dot{\rho}_{de}}{3H\rho_{dm}},\label{e-rdm}
\end{equation}
\begin{equation}
\gamma_{de}^{e}=\gamma_{de}+\gamma_{dm}r+\frac{\dot{\rho}_{dm}}{3H\rho_{de}},\label{e-rde}
\end{equation}
where $r=\rho_{dm}/\rho_{de}$ and
$\gamma_{i}=\frac{p_{i}}{\rho_{i}}+1$. When
$\gamma_{dm}^{e}=\gamma_{dm}$ and $\gamma_{de}^{e}=\gamma_{de}$,
Eqs.(\ref{C-dm}) and (\ref{C-de}) are reduced to the non-interacting
cases. In addition, for the introduced effective barotropic indexes
$\gamma_{i}^{e}$ and the parameter $r$, they have the relations:
\begin{equation}
(\gamma_{dm}^{e}-\gamma_{dm})r+(\gamma_{de}^{e}-\gamma_{de})=0,\label{relation1}
\end{equation}
\begin{equation}
\dot{r}=-3Hr(\gamma_{dm}^{e}-\gamma_{de}^{e}).\label{relation2}
\end{equation}
Considering that the effective barotropic index of dark energy is
 given by $\gamma_{de}^{e}=\gamma_{de}-F(r)$  \cite{Fr} with $F(r)$ being a function of the energy density ratio $r$,
 one get,
\begin{equation}
\gamma_{dm}^{e}-\gamma_{de}^{e}=\gamma_{dm}-\gamma_{de}+F(r)(1+\frac{1}{r}),\label{det}
\end{equation}
 and the energy conservation equations (\ref{C-dm}) and (\ref{C-de})
 become
\begin{equation}
\dot{\rho}_{dm}+3H\rho_{dm}\gamma_{dm}=-3H\rho_{de}F(r),\label{conservation1}
\end{equation}
\begin{equation}
\dot{\rho}_{de}+3H\rho_{de}\gamma_{de}=3H\rho_{de}F(r).\label{conservation2}
\end{equation}
It is obvious that $F(r)$ can be dubbed as interaction function,
which measures the strength of interaction. From the Eqs.
(\ref{conservation1}) and (\ref{conservation2}), we can see the
energy transfer between dark energy and dark matter, and for
$F(r)=0$  Eqs. (\ref{conservation1}) and (\ref{conservation2})
reduce to the non-interacting cases for the energy conservation
equation. In the following we consider two concrete forms of
interaction function $F(r)$.

\subsection{$\text{Interacting dark model with a  constant interaction function F(r)}$}

For calculation, following Ref. \cite{Fr} we consider a concrete
form of the constant function $F(r)$ as
\begin{equation}
F(r)=-\frac{r_{\infty}}{1+r_{\infty}}(\gamma_{dm}-\gamma_{de}),\label{F-r}
\end{equation}
where $\gamma_{dm}$ and  $\gamma_{de}$ are considered as constant,
and the parameter $r_{\infty}$ is also a constant which denotes the
ratio between the energy densities at infinity. From Eq. (\ref{F-r})
it is easy to see that for the parameter $r_{\infty}=0$, the
non-interacting energy conservation equations for dark matter and
dark energy are obtained in Eqs. (\ref{conservation1}) and
(\ref{conservation2}). Integrating Eqs. (\ref{relation2}),
(\ref{C-dm}), and (\ref{C-de}), we get the expressions of the energy
density for dark matter and dark energy,
\begin{equation}
\rho_{dm}=\rho_{0de}[r_{\infty}+(r_{0}-r_{\infty})(1+z)^{3\alpha}](1+z)^{\beta},\label{rho1}
\end{equation}
\begin{equation}
\rho_{de}=\rho_{0de}(1+z)^{\beta},\label{rho2}
\end{equation}
where $r_{0}$ denotes the current value of the parameter $r$, and
two defined parameter,
\begin{equation}
\alpha=\frac{\gamma_{dm}-\gamma_{de}}{1+r_{\infty}},\label{alpha}
\end{equation}
\begin{equation}
\beta=3\frac{r_{\infty}\gamma_{dm}+\gamma_{de}}{1+r_{\infty}}.\label{beta}
\end{equation}
Considering the definitions of the dimensionless energy densities,
$\Omega_{0b}=\frac{8 \pi G \rho_{0b}}{3H_{0}^{2}}$,
$\Omega_{0dm}=\frac{8 \pi G \rho_{0dm}}{3H_{0}^{2}}$ and
$\Omega_{0de}=\frac{8 \pi G \rho_{0de}}{3H_{0}^{2}}$,  the Friedmann
equation can be written as
\begin{eqnarray}
E^{2}=\frac{H^{2}}{H_{0}^{2}} &&{=}\frac{8 \pi
G}{3H_{0}^{2}}(\rho_{b}+\rho_{dm}+\rho_{de})\nonumber\\
&&{=}\Omega_{0b}(1+z)^{3}+(1-\Omega_{0b}-\Omega_{0de}-r_{\infty}\Omega_{0de})(1+z)^{3}
+\Omega_{0de}(1+r_{\infty})(1+z)^{\beta}\nonumber\\
&&{=}\Omega_{0m}(1+z)^{3}+(1-\Omega_{0m})(1+r_{\infty})(1+z)^{\beta}-(1-\Omega_{0m})r_{\infty}(1+z)^{3},\label{H-E}
\end{eqnarray}
with using the relation $\Omega_{0b}+\Omega_{0dm}+\Omega_{0de}=1$.

\subsection{$\text{Interacting dark model with a  variable interaction function F(r)}$}
In this part we consider a concrete variable interaction function
$F(r)$ to investigate the interaction between dark matter and dark
energy. Following Ref. \cite{Fr}, one possible choice for the
function $F(r)$ is
\begin{equation}
F(r)=-\frac{(1-r)r_{\infty}^{2}}{r(1-r_{\infty}^{2})}(\gamma_{dm}-\gamma_{de}).\label{Fr-variable}
\end{equation}
With using Eq. (\ref{Fr-variable}), integrating Eqs. (\ref{C-dm})
and (\ref{C-de}) one can get the energy densities of dark matter and
dark energy
\begin{equation}
\rho_{dm}=\rho_{0de}\sqrt{r_{\infty}^{2}+(r_{0}^{2}-r_{\infty}^{2})(1+z)^{\mu}}
(1+z)^{3\nu}[\frac{(1-r/r_{\infty})(1+r_{0}/r_{\infty})}{(1-r_{0}/r_{\infty})
(1+r/r_{\infty})}]^{\frac{r_{\infty}}{2}},\label{rhodm-Fv}
\end{equation}
\begin{equation}
\rho_{de}=\rho_{0de}(1+z)^{3\nu}
[\frac{(1-r/r_{\infty})(1+r_{0}/r_{\infty})}{(1-r_{0}/r_{\infty})(1+r/r_{\infty})}]^{\frac{r_{\infty}}{2}},\label{rhode-Fv}
\end{equation}
where
\begin{equation}
\mu=\frac{6(\gamma_{dm}-\gamma_{de})}{1-r_{\infty}^{2}},
~~\nu=\gamma_{de}-\frac{(\gamma_{dm}-\gamma_{de})r_{\infty}^{2}}{1-r_{\infty}^{2}}.\label{munu}
\end{equation}
So, the expression of dimensionless Hubble parameter can be written
as
\begin{equation}
E(z)^{2}=\frac{H(z)^{2}}{H^{2}_{0}}=\Omega_{0de}(1+z)^{3\nu}
[\frac{(1-r/r_{\infty})(1+r_{0}/r_{\infty})}{(1-r_{0}/r_{\infty})(1+r/r_{\infty})}]^{\frac{r_{\infty}}{2}}
[1+\sqrt{r_{\infty}^{2}+(r_{0}^{2}-r_{\infty}^{2})(1+z)^{\mu}}]+\Omega_{0b}(1+z)^{3}.\label{H-E-Fv}
\end{equation}
For above two interacting cases, according to  Eqs. (\ref{H-E}) and
(\ref{H-E-Fv}) one can see that they are reduced to the
non-interacting case with a model-independent dark energy scenario
$w=w_{0}$=constant, when the parameter $r_{\infty}=0$.


\section{$\text{Cosmological constraints on the interacting models of  dark sectors}$}\label{constraint}

In the following we apply the current observational data to
constrain the above interacting models of  dark matter and dark
energy. For the used observational data, we consider 557 Union2
dataset of type supernovae Ia (SNIa) \cite{557Union2}, observational
Hubble data (OHD) \cite{OHD}, X-ray gas mass fraction in  cluster
\cite{ref:07060033}, baryon acoustic oscillation (BAO)
\cite{ref:Percival2}, and cosmic microwave background (CMB) data
\cite{7ywmap}.

\subsection{Type Ia supernovae}

For SNIa observation,  distance modulus $\mu(z)$ is expressed as
\begin{equation}
\mu_{th}(z)=5\log_{10}[D_{L}(z)]+\mu_{0},
\end{equation}
where $D_{L}(z)=H_0 d_L(z)/c$ is the Hubble-free luminosity
distance, with $H_0$  being the Hubble constant defined by the
re-normalized quantity $h$ as $H_0 =100 h~{\rm km ~s}^{-1} {\rm
Mpc}^{-1}$,
 and
\begin{eqnarray}
d_L(z)&=&c(1+z) \int_0^z\frac{dz'}{H(z')}, \nonumber\\
 &&\mu_{0}=5log_{10}(\frac{H_{0}^{-1}}{Mpc})+25=42.38-5log_{10}h,\nonumber
\end{eqnarray}
for a flat-geometry universe.
 Additionally, the observed distance moduli $\mu_{obs}(z_i)$ of SNIa at $z_i$ is
\begin{equation}
\mu_{obs}(z_i) = m_{obs}(z_i)-M,
\end{equation}
where $M$ is their absolute magnitudes.

For using SNIa data, theoretical model parameters  $\theta$ can be
determined by a likelihood analysis, based on the calculation of
\begin{eqnarray}
\chi^2(\theta,M^{\prime})\equiv \sum_{SNIa}\frac{\left\{
\mu_{obs}(z_i)-\mu_{th}(\theta,z_i)\right\}^2} {\sigma_i^2}
=\sum_{SNIa}\frac{\left\{ 5 \log_{10}[D_L(\theta,z_i)] -
m_{obs}(z_i) + M^{\prime} \right\}^2} {\sigma_i^2}, \ \ \ \
\label{eq:chi2}
\end{eqnarray}
where $M^{\prime}\equiv\mu_0+M$ is a nuisance parameter which
includes the absolute magnitude and the parameter $h$. The nuisance
  parameter $M^{\prime}$ can be marginalized over
analytically
\cite{ref:SNchi2,ref:SNchi21,ref:SNchi22,ref:SNchi23,ref:SNchi24,ref:SNchi25,ref:SNchi26}
as
\begin{equation}
\bar{\chi}^2(\theta) = -2 \ln \int_{-\infty}^{+\infty}\exp \left[
-\frac{1}{2} \chi^2(\theta,M^{\prime}) \right] dM^{\prime},\nonumber
\label{eq:chi2marg}
\end{equation}
resulting to
\begin{equation}
\bar{\chi}^2 =  A - \frac{B^2}{C} + \ln \left( \frac{C}{2\pi}\right)
, \label{eq:chi2mar}
\end{equation}
with
\begin{eqnarray}
&&A=\sum_{SNIa} \frac {\left\{5\log_{10}
[D_L(\theta,z_i)]-m_{obs}(z_i)\right\}^2}{\sigma_i^2},\nonumber\\
&& B=\sum_{SNIa} \frac {5
\log_{10}[D_L(\theta,z_i)]-m_{obs}(z_i)}{\sigma_i^2},\nonumber
\\
&& C=\sum_{SNIa} \frac {1}{\sigma_i^2}\nonumber.
\end{eqnarray}
Noting that the expression
\begin{equation}
\chi^2_{SNIa}(\theta)=A-(B^2/C),\label{eq:chi2SN}\nonumber
\end{equation}
which is equivalent  to (\ref{eq:chi2mar}) except a constant, then
it is often used in the likelihood analysis, since in this case the
constraint results will not be affected by the nuisance parameter
$M^{\prime}$.

\subsection{Observational Hubble data}

The observational Hubble data \cite{ohdzhang} are based on
differential ages of the galaxies. In \cite{ref:JVS2003}, Jimenez
{\it et al.} obtained an independent estimate for the Hubble
parameter using the method developed in \cite{ref:JL2002}, and used
it to constrain the cosmological models. The Hubble parameter
depending on the differential ages as a function of redshift $z$ can
be written in the form of
\begin{equation}
H(z)=-\frac{1}{1+z}\frac{dz}{dt}.
\end{equation}
So, once $dz/dt$ is known, $H(z)$ is obtained directly. By using the
differential ages of passively-evolving galaxies from the Gemini
Deep Deep Survey (GDDS) \cite{ref:GDDS} and archival data
\cite{ref:archive6}, Simon {\it et al.} obtained several values of
$H(z)$  at  different redshift \cite{OHD}. The twelve observational
Hubble data  (redshift interval  $0\lesssim z \lesssim 1.8$) from
\cite{12Hubbledata,12Hubbledata1,H0prior,3Hubbledata} are listted in
Table \ref{table-12Hubbledata}.
\begin{table}[ht]
\begin{center}
\begin{tabular}{c|llllllllllll}
\hline\hline
 $z$ &\ 0 & 0.1 & 0.17 & 0.27 & 0.4 & 0.48 & 0.88 & 0.9 & 1.30 & 1.43 & 1.53 & 1.75  \\ \hline
 $H(z)\ ({\rm km~s^{-1}\,Mpc^{-1})}$ &\ 74.2 & 69 & 83 & 77 & 95 & 97 & 90 & 117 & 168 & 177 & 140 & 202  \\ \hline
 $1 \sigma$ uncertainty &\ $\pm 3.6$ & $\pm 12$ & $\pm 8$ & $\pm 14$ & $\pm 17$ & $\pm 60$ & $\pm 40$
 & $\pm 23$ & $\pm 17$ & $\pm 18$ & $\pm 14$ & $\pm 40$ \\
\hline\hline
\end{tabular}
\end{center}
\caption{\label{table-12Hubbledata} The observational $H(z)$
data~\cite{12Hubbledata,12Hubbledata1,H0prior,3Hubbledata}.}
\end{table}
In addition, in \cite{3Hubbledata} the authors take the BAO scale as
a standard ruler in the radial direction, and obtain three more
additional data: $H(z=0.24)=79.69\pm2.32, H(z=0.34)=83.8\pm2.96,$
and $H(z=0.43)=86.45\pm3.27$.

 The best fit values of the model parameters from observational Hubble data  are determined by minimizing \cite{chi2hub,chi2hub1,chi2hub2}
 \begin{equation}
 \chi_{OHD}^2(H_{0},\theta)=\sum_{i=1}^{15} \frac{[H_{th}(H_{0},\theta;z_i)-H_{obs}(z_i)]^2}{\sigma^2(z_i)},\label{chi2OHD}
 \end{equation}
 where  $H_{th}$ is the predicted value for the Hubble parameter, $H_{obs}$ is the observed value, $\sigma(z_i)$ is the standard
 deviation measurement uncertainty, and the summation is over the $15$ observational Hubble data points at redshifts $z_i$.

\subsection{The X-ray gas mass fraction}

The observations of X-ray gas mass fraction in galaxy clusters
provide the information on the dark matter and the formation of
structure, so they can be used to constrain the cosmological
parameters. It is assumed that the baryon gas mass fraction in
clusters \cite{30Nesseris}
\begin{equation}
f_{gas}=\frac{M_{b-gas}}{M_{tot}} \label{11f-gas}
\end{equation}
is constant, independent of redshift and is related to the global
fraction of the universe $\Omega_{b}/\Omega_{0m}$. In the standard
cold dark matter (SCDM) model, $f_{gas}^{SCDM}$ is \cite{30Nesseris}
\begin{equation}
f_{gas}^{SCDM}=\frac{b}{1+\alpha}\frac{\Omega_{b}}{\Omega_{0m}}(\frac{d_{A}^{SCDM}(z)}{d_{A}(z)})^{\frac{3}{2}},\label{12f-gasSCDM}
\end{equation}
where $d_{A}$ is diameter distance which relates with $d_{L}$ via
$d_{L}(z)=(1+z)^{2}d_{A}(z)$, the parameter $b$ is a bias factor
 suggesting that the baryon fraction in clusters is slightly lower than
for the universe as a whole, the parameter
$\alpha\simeq0.19\sqrt{h}$ is the ratio factor of optically luminous
baryonic mass with X-ray gas contained in clusters. From Cluster
Baryon Fraction (CBF), the best fit values of parameters in
cosmological model  can be determined by minimizing
\cite{30Nesseris}
\begin{equation}
\chi^{2}_{CBF}(\theta)=C-\frac{B^{2}}{A},\label{chi2-CBF}
\end{equation}
where
\begin{eqnarray*}
A=\sum_{i=1}^{N}\frac{\widetilde{f}_{gas}^{SCDM}(z_{i})^{2}}{\sigma^{2}_{f_{gas,i}}},
\end{eqnarray*}

\begin{eqnarray*}
B=\sum_{i=1}^{N}\frac{\widetilde{f}_{gas}^{SCDM}(z_{i})\cdot
f_{gas,i}}{\sigma^{2}_{f_{gas,i}}},
\end{eqnarray*}

\begin{equation}
C=\sum_{i=1}^{N}\frac{f_{gas,i}^{2}}{\sigma^{2}_{f_{gas,i}}},\label{14ABC}
\end{equation}

and
\begin{equation}
\widetilde{f}_{gas}^{SCDM}(z_{i})=(\frac{d_{A}^{SCDM}(z)}{d_{A}(z)})^{\frac{3}{2}}.\label{15fgas}
\end{equation}
$N=42$ is the number of the observed $f_{gas,i}$ and
$\sigma^{2}_{gas,i}$ published in Ref. \cite{42Allen}.

\subsection{Baryon acoustic oscillation}

The baryon acoustic oscillations are detected in the clustering of
the 2dFGRS  and   SDSS main galaxy samples, which measure the
distance-redshift relation. The value of dimensionless parameter $A$
can be calculated from these samples, which is defined by
\begin{equation}
A=\sqrt{\Omega_{0m}}E(z_{BAO})^{-1/3}[\frac{1}{z_{BAO}}\int_{0}^{z}\frac{dz^{'}}{E(z^{'};\theta)}]^{2/3},\label{9A-BAO}
\end{equation}
where $E(z)$ is included in the Hubble parameter $H(z)=H_{0}E(z)$,
and the values of $z_{BAO}=0.35$ and $A=0.469\pm0.017$ are given by
measuring  from the SDSS \cite{BAOdata-A1,BAOdata-A2,BAOdata-A3}.
One can minimize the $\chi^{2}_{BAO}$ defined as
\begin{equation}
\chi^{2}_{BAO}(\theta)=\frac{(A(\theta)-0.469)^{2}}{0.017^{2}}.\label{chi2-BAO}
\end{equation}

\subsection{Cosmic microwave background}

For CMB data, we use the CMB shift parameter $R$ to constrain the
cosmological model. It is defined by \cite{ref:Bond1997}
\begin{equation}
 R=\sqrt{\Omega_{0m} H^2_0}(1+z_{\ast})D_A(z_{\ast})/c=\sqrt{\Omega_{m}}\int_{0}^{z_{\ast}}\frac{H_{0}dz^{'}}{H(z^{'};\theta)},\label{R-CMB}
\end{equation}
here  $z_{\ast}$ is the redshift at the decoupling epoch of photons,
which is obtained from the 7yWMAP data  $z_{\ast}=1091.3$, and the
value of $R$ is given by \cite{7ywmap}
\begin{equation}
R=1.725\pm0.018.\label{7R-value}
\end{equation}
From the CMB constraint, the best fit values of parameters in the DE
models can be determined by minimizing
\begin{equation}
\chi^{2}_{CMB}(\theta)=\frac{(R(\theta)-1.725)^{2}}{0.018^{2}}.\label{chi2-CMB}
\end{equation}

 The total $\chi^{2}$ is expressed as
 \begin{equation}
 \chi^{2}_{total}(\theta)=\sum_{i}\chi^{2}_{i}(\theta),\label{chi2total-5data}
 \end{equation}
here $\theta$ denotes the model parameters, and suffix $i$ denotes
any one observational data of above five data: SNIa, OHD, CBF, BAO
and CMB. In this expression,
 for each observation $\chi^{2}$ corresponds to
Eqs.(\ref{eq:chi2SN}), (\ref{chi2OHD}), (\ref{chi2-CBF}),
(\ref{chi2-BAO}) and (\ref{chi2-CMB}), respectively.
 Using the currently observed data,
  Fig. \ref{figure-ab-Fr-C}, Fig. \ref{figure-ab-Fr-V} and Fig. \ref{figure-ab-nonI-w=c}
   respectively plot the 2-D contours with  $1\sigma, 2\sigma$
  confidence levels of model parameters  in the flat universe for the case of $F(r)=$constant, $F(r)=$variable and
  non-interacting model of $w=w_{0}=$constant.
  And for each model  we consider three different combined constraints on model parameters,
  i.e. respectively using the combined data of SNIa+OHD+BAO, SNIa+CMB and SNIa+OHD+CBF+BAO+CMB.  The
   corresponding calculation results for the constraints on model parameters are listed
   in table \ref{table-ab-Fr-c}, \ref{table-ab-Fr-v}  and \ref{table-ab-nonI-w=w0}. According to
   these three constraints on the parameter $r_{\infty}$, as shown in table
   \ref{table-ab-Fr-c} one can see that for the case of the constant interaction
   function $F(r)$,  there exist a
   non-vanishing, but weak interaction. However,  for the case of the
   variable interaction function $F(r)$, considering that the best fit values of $r_{\infty}$ are near
  to zero  it seems that the
 observational data tends to have no interaction between dark matter
 and dark energy, but the confidence levels of this parameter are still
 wide. Also, from table \ref{table-ab-Fr-c}, \ref{table-ab-Fr-v}  and \ref{table-ab-nonI-w=w0}
  it can be seen that the most stringent constraint on model parameters is
 given by using the most observational data: SNIa+OHD+CBF+BAO+CMB, when compare three combined constraints.
 In addition, by using the best fit values of model parameters we
 can obtain the values of state parameter for dark energy
 $w_{de}$, according to the calculation formula
 $w_{de}=\gamma_{de}-1=\frac{\beta(1+r_{\infty})}{3}-r_{\infty}\gamma_{dm}-1$
 for the case of constant interaction function, and
  $w_{de}=\gamma_{de}-1$ for the case of variable interaction function.
 It is shown that for both interacting scenarios,   the values of state
 parameter $w_{de}$
 are  in  phantom region ($w_{de}<-1$) for the combined constraint from
 SNIa+OHD+BAO data, and are in quintessence region ($w_{de}>-1$) for the combined constraint from
 SNIa+OHD+CBF+BAO+CMB data, which are consistent with the constraint
 results of the non-interacting case.


\begin{figure}[ht]
  \includegraphics[width=4cm]{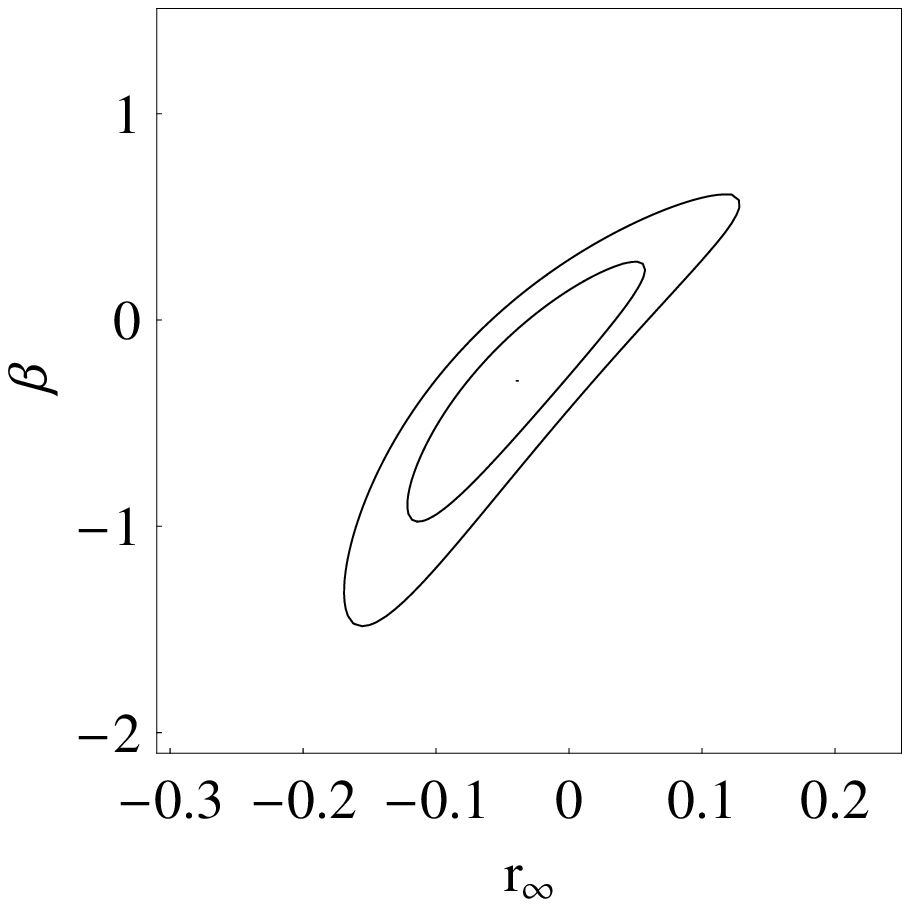}~
  \includegraphics[width=4cm]{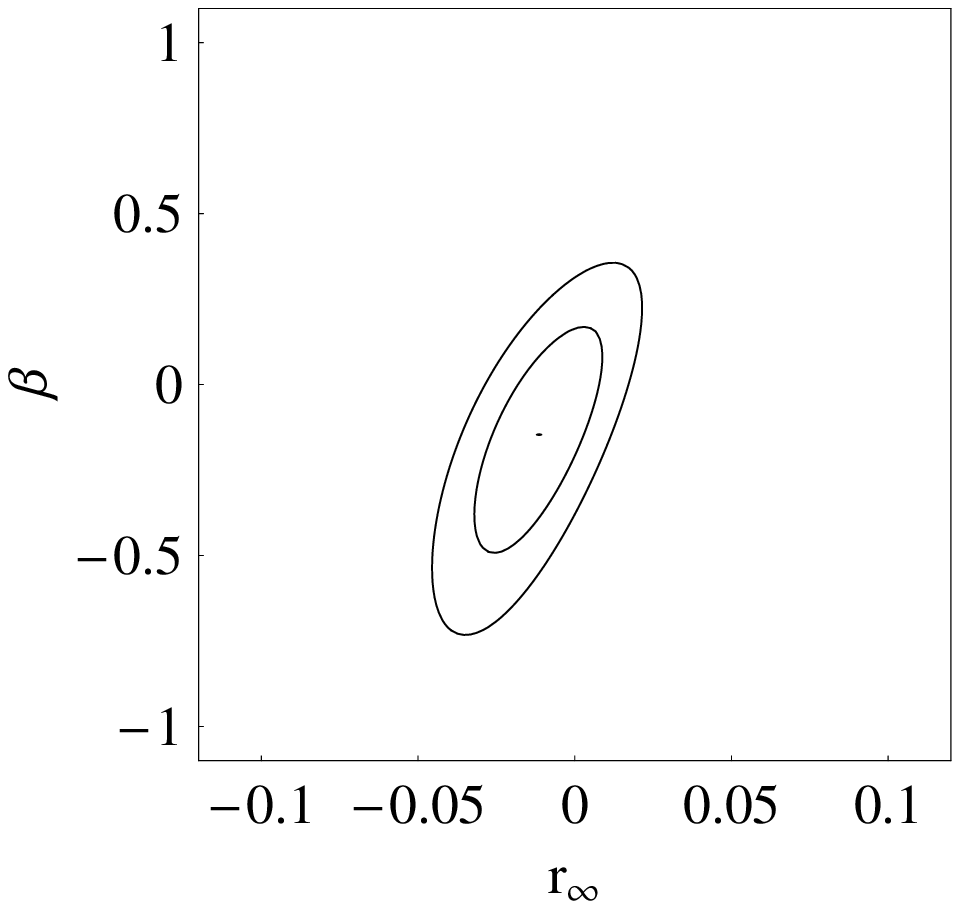}
  \includegraphics[width=4cm]{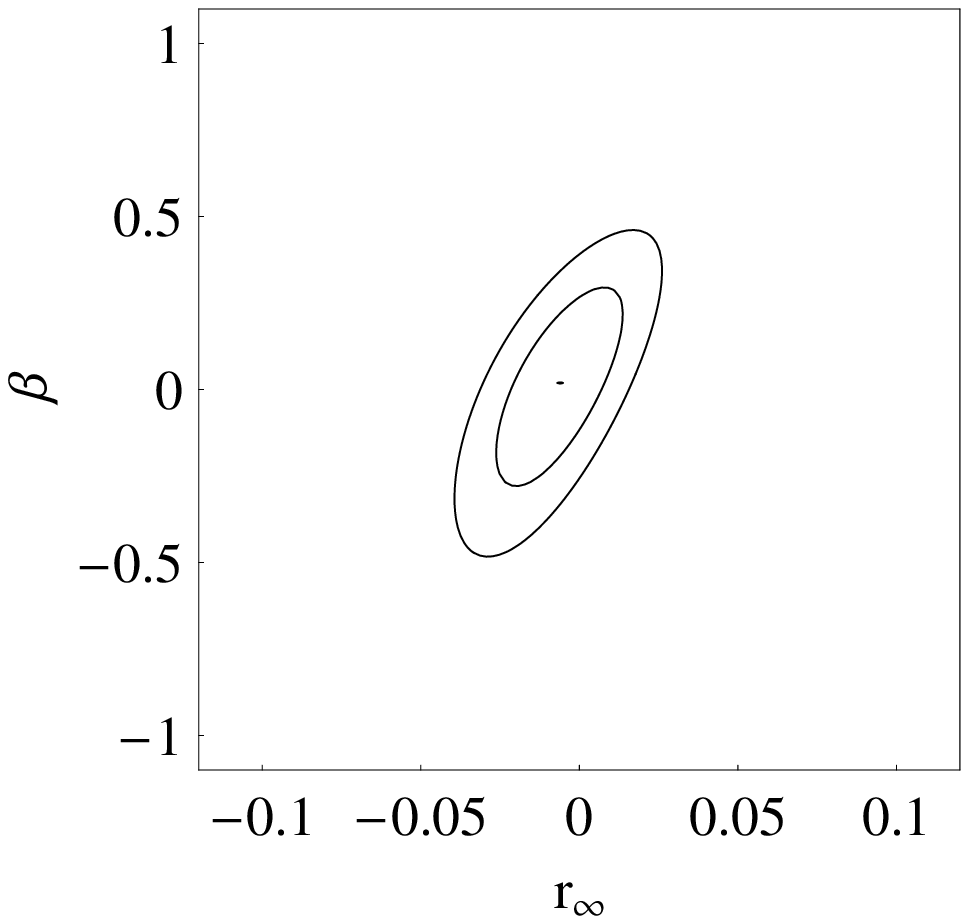}~\\
   \includegraphics[width=4cm]{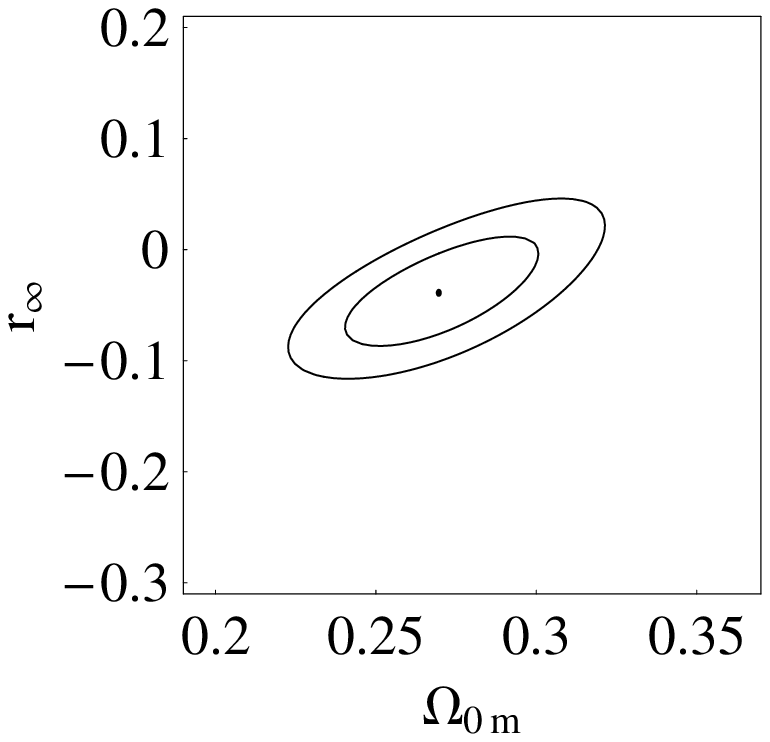}~
  \includegraphics[width=4cm]{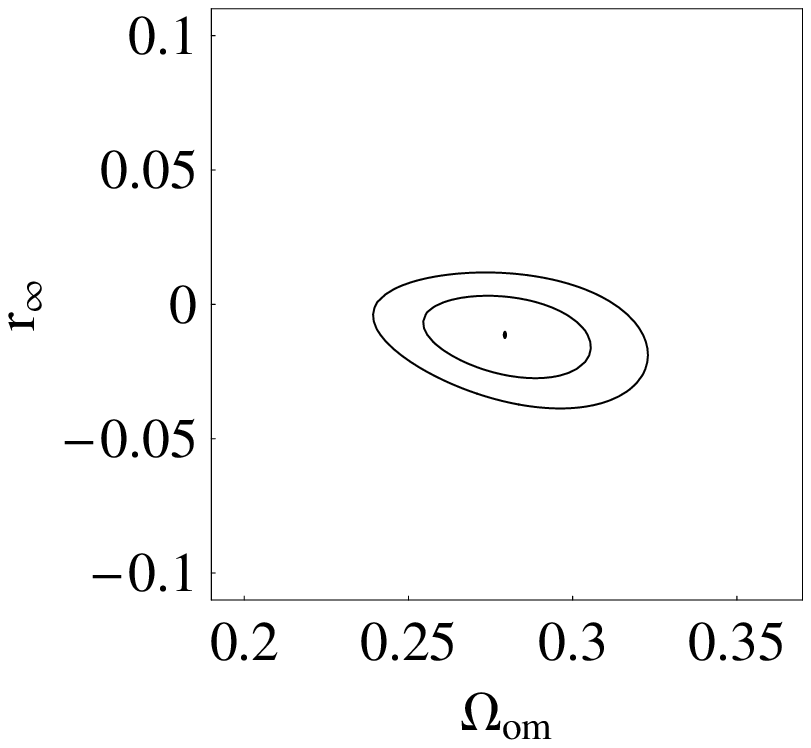}
  \includegraphics[width=4cm]{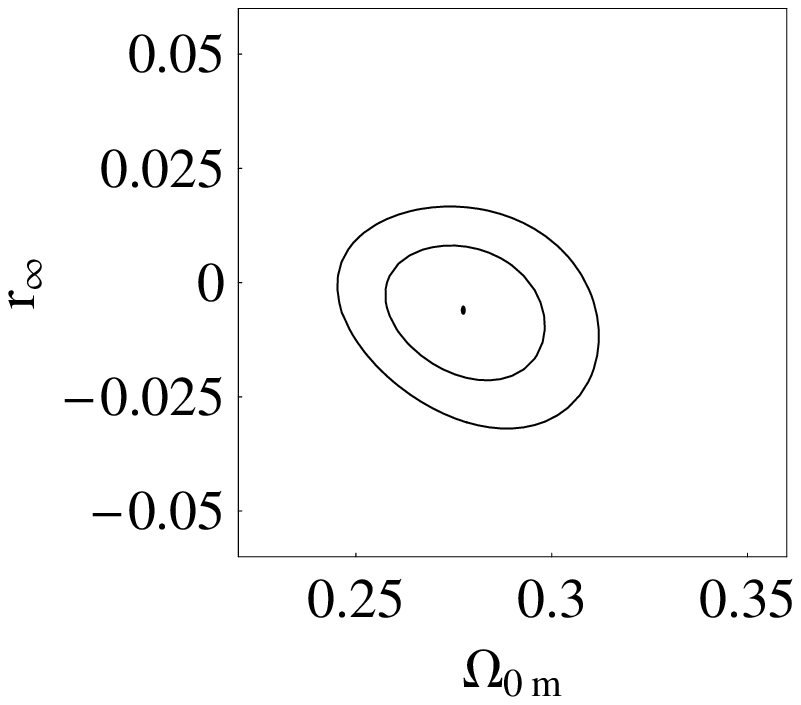}~\\
  \caption{The 2-D contours with  $1\sigma$ and $2\sigma$ confidence levels of model parameters
   in the interacting model with owing a  constant function $F(r)$
    from the current observational data: SNIa+OHD+BAO (left), SNIa+CMB (middle) and
   SNIa+OHD+CBF+BAO+CMB  (right).}\label{figure-ab-Fr-C}
\end{figure}
\begin{figure}[ht]
  \includegraphics[width=4cm]{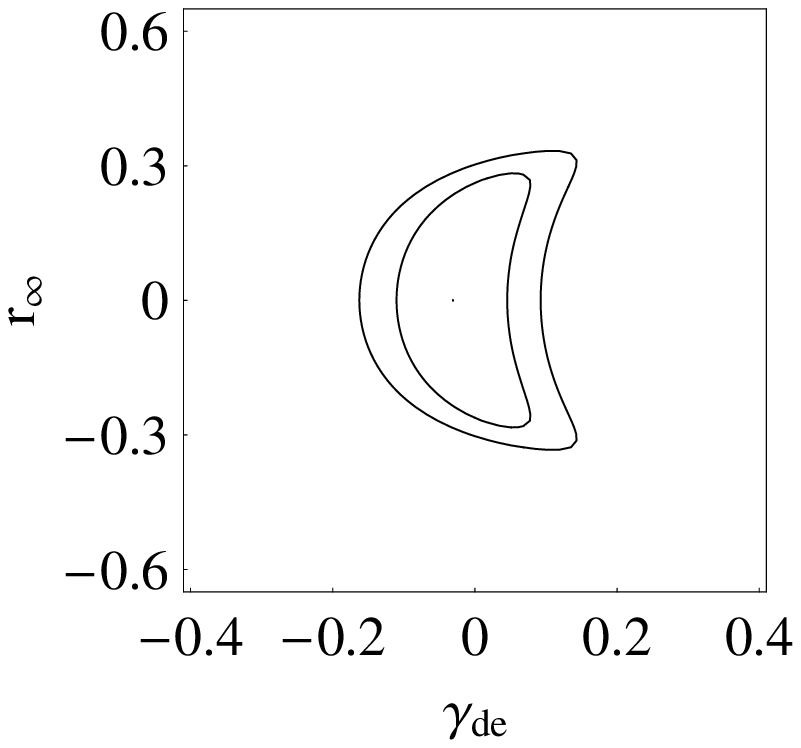}~
  \includegraphics[width=4cm]{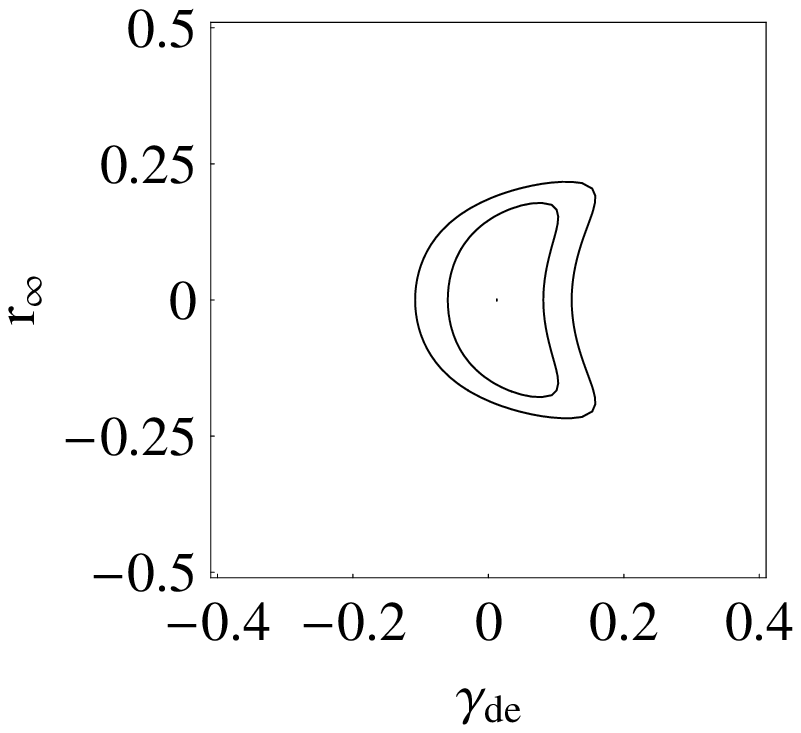}~
  \includegraphics[width=4cm]{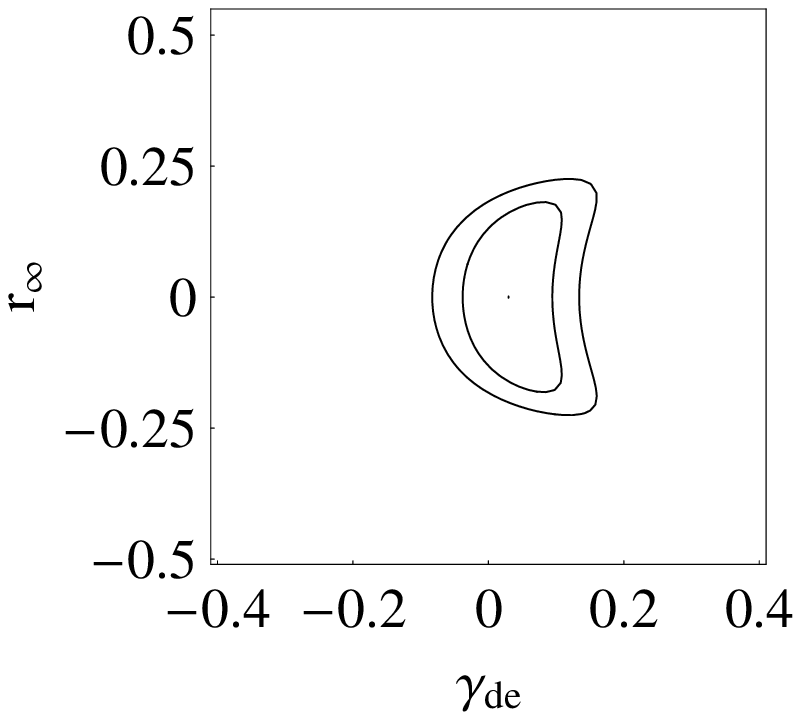}~\\
  \includegraphics[width=4cm]{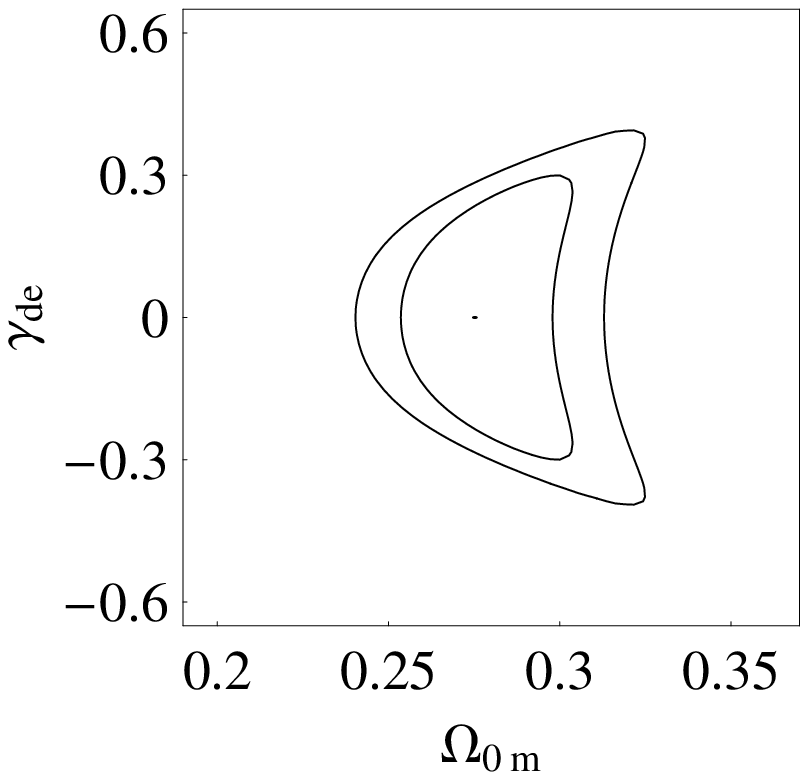}~
  \includegraphics[width=4cm]{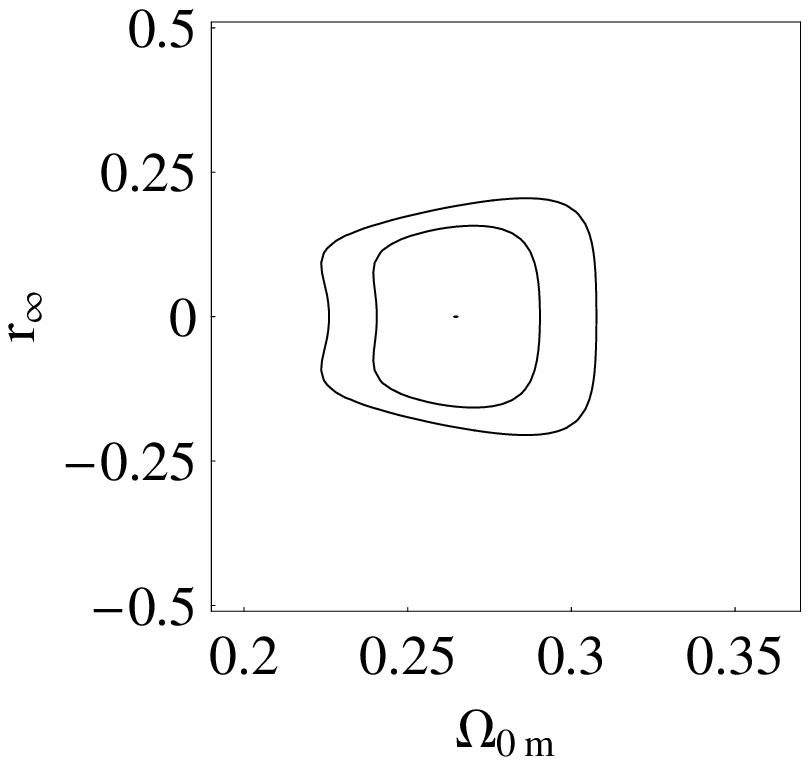}~
  \includegraphics[width=4cm]{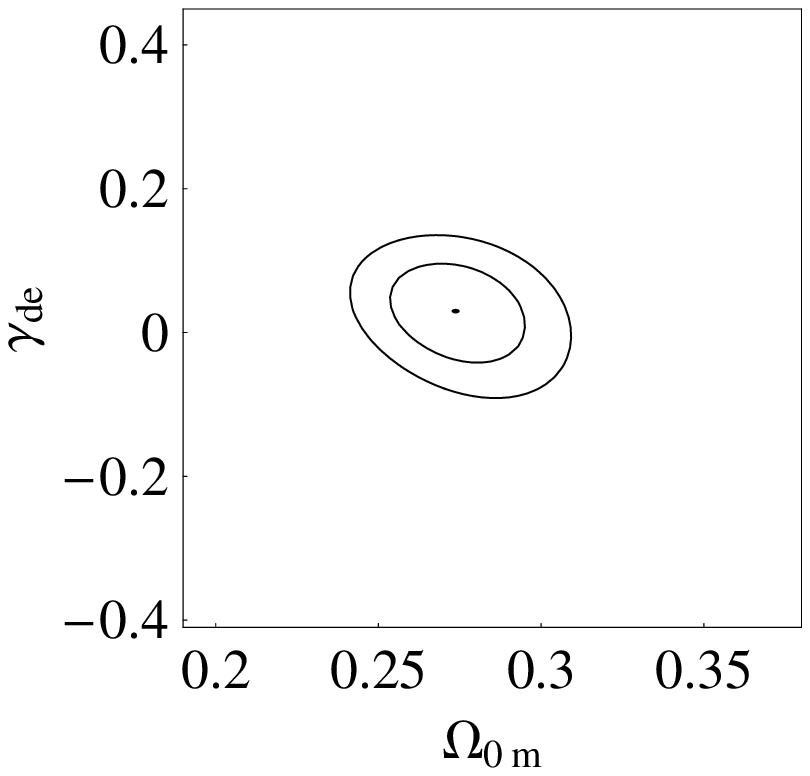}~\\
  \caption{The 2-D contours with  $1\sigma$ and $2\sigma$ confidence levels of model parameters
   in the interacting model with owing a variable function $F(r)$
    by using  SNIa+OHD+BAO (left), SNIa+CMB (middle) and
   SNIa+OHD+CBF+BAO+CMB data (right).}\label{figure-ab-Fr-V}
\end{figure}
\begin{figure}[ht]
  \includegraphics[width=4cm]{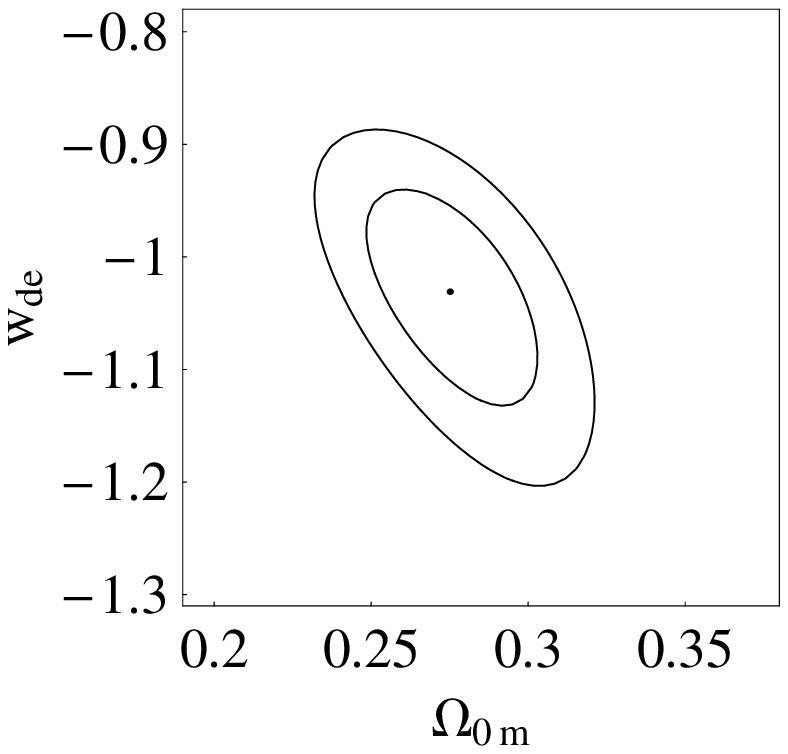}~
  \includegraphics[width=4cm]{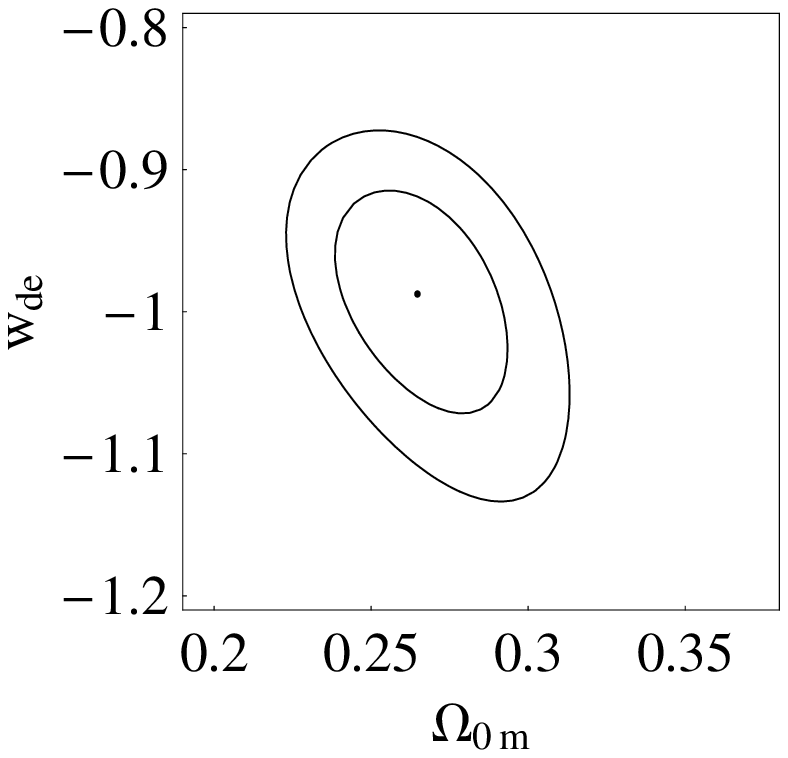}~
  \includegraphics[width=4cm]{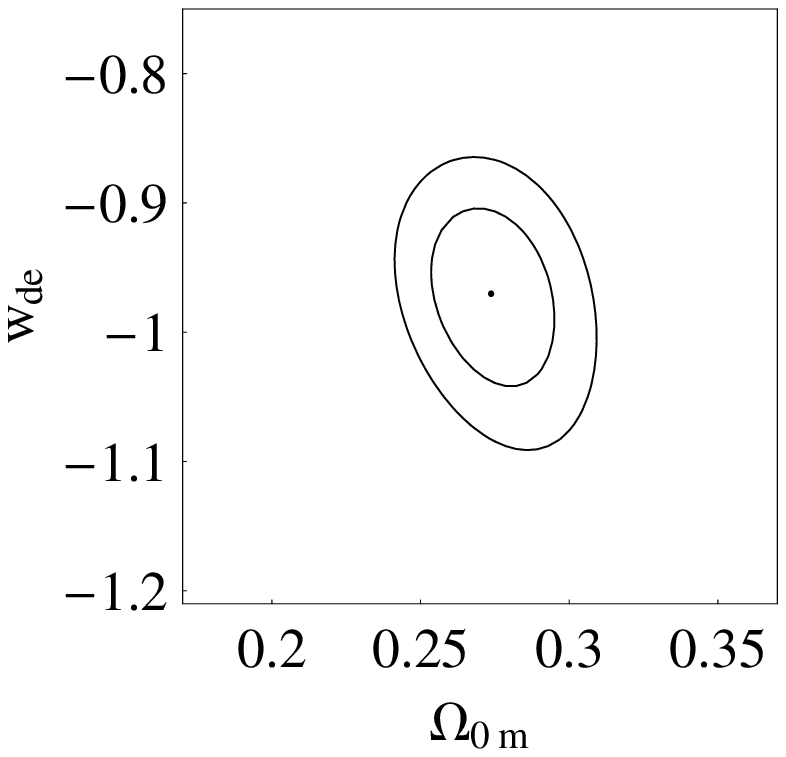}\\
  \caption{The 2-D contours with  $1\sigma$ and $2\sigma$ confidence levels of model parameters
   in the non-interacting model of $w=w_{0}$=constant by using SNIa+OHD+BAO (left), SNIa+CMB (middle) and
   SNIa+OHD+CBF+BAO+CMB data (right).}\label{figure-ab-nonI-w=c}
\end{figure}
\begin{table}[ht]
 \vspace*{-12pt}
 \begin{center}
 \begin{tabular}{c |c |c |  c|  c|  c} \hline\hline
            & $\chi_{min}^{2}$ & $\chi_{min}^{2}/dof$ & $\Omega_{0m}$ & $r_{\infty}$ &  $\beta$    \\\hline
 SNIa+OHD+BAO & 554.092& 0.967 & $0.270^{+0.033+0.053}_{-0.032-0.049}$
            & $-0.039^{+0.096+0.167}_{-0.083-0.122}$
            & $-0.295^{+0.577+0.904}_{-0.682-1.189}$\\\hline
 SNIa+CMB   & 542.633 & 0.972 & $0.279^{+0.029+0.046}_{-0.026-0.042}$
            & $-0.011^{+0.020+0.032}_{-0.021-0.033}$
            & $-0.146^{+0.315+0.502}_{-0.346-0.586}$\\\hline
 SNIa+OHD+CBF+BAO+CMB & 616.397& 1.001& $0.277^{+0.023+0.037}_{-0.021-0.034}$
                        & $-0.006^{+0.020+0.032}_{-0.020-0.033}$
                        & $0.019^{+0.275+0.442}_{-0.298-0.492}$ \\\hline\hline
 \end{tabular}
 \end{center}
 \caption{  The values of $\chi_{min}^{2}$, $\chi_{min}^{2}/dof$, and
 the best fit values of model parameters with their confidence levels for the constant interacting model  from
 the current observational data: SNIa+OHD+BAO, SNIa+CMB and SNIa+OHD+CBF+BAO+CMB,
 where the value of $dof$ (degree of freedom)
 equals the number of observational data points minus the number of
 model parameters. }\label{table-ab-Fr-c}
 \end{table}
\begin{table}[ht]
 \vspace*{-12pt}
 \begin{center}
 \begin{tabular}{c |c |c |  c|  c|  c} \hline\hline
            & $\chi_{min}^{2}$ & $\chi_{min}^{2}/dof$&
             $\Omega_{0m}$  & $r_{\infty}$ &  $\gamma_{de}$    \\\hline
 SNIa+OHD+BAO & 554.41& 0.968& $0.275^{+0.031+0.052}_{-0.023-0.037}$
            & $-0.00001^{+0.586+0.919}_{-0.692-1.211}$
            & $-0.031^{+0.091+0.158}_{-0.077-0.121}$\\\hline
 SNIa+CMB   & 542.73 & 0.973& $0.265^{+0.027+0.045}_{-0.028-0.043}$
            & $-0.0005^{+0.1788+0.2175}_{-0.1778-0.2165}$
            & $0.012^{+0.091+0.146}_{-0.072-0.120}$\\\hline
 SNIa+OHD+CBF+BAO+CMB & 616.537& 1.001& $0.274^{+0.023+0.037}_{-0.022-0.035}$
            & $-0.00004^{+0.18147+0.22541}_{-0.18139-0.22526}$
            & $0.030^{+0.078+0.130}_{-0.068-0.113}$\\\hline\hline
 \end{tabular}
 \end{center}
 \caption{  The values of $\chi_{min}^{2}$, $\chi_{min}^{2}/dof$, and
 the best fit values of model parameters with their confidence
 levels   for the variable interacting model from
 the current observational data.}\label{table-ab-Fr-v}
 \end{table}
\begin{table}[ht]
 \vspace*{-12pt}
 \begin{center}
 \begin{tabular}{c |c |c |  c|  c} \hline\hline
            & $\chi_{min}^{2}$ & $\chi_{min}^{2}/dof$
            & $\Omega_{0m}$ &  $w_{0}$    \\\hline
 SNIa+OHD+BAO & 554.410 & 0.968
            & $0.275^{+0.028+0.046}_{-0.027-0.043}$
            & $-1.031^{+0.091+0.144}_{-0.101-0.172}$\\\hline
 SNIa+CMB   & 542.730 & 0.973
            & $0.265^{+0.028+0.048}_{-0.026-0.042}$
            & $-0.988^{+0.073+0.116}_{-0.084-0.145}$\\\hline
 SNIa+OHD+CBF+BAO+CMB & 616.537 & 1.001
                        & $0.274^{+0.021+0.0.035}_{-0.020-0.033}$
                        & $-0.970^{+0.066+0.106}_{-0.072-0.121}$ \\\hline\hline
 \end{tabular}
 \end{center}
 \caption{  The values of $\chi_{min}^{2}$, $\chi_{min}^{2}/dof$, and
 the best fit values of model parameters with their confidence levels for the non-interacting model  from
 the current observational data.}\label{table-ab-nonI-w=w0}
 \end{table}

 \begin{figure}[ht]
   \includegraphics[width=5cm]{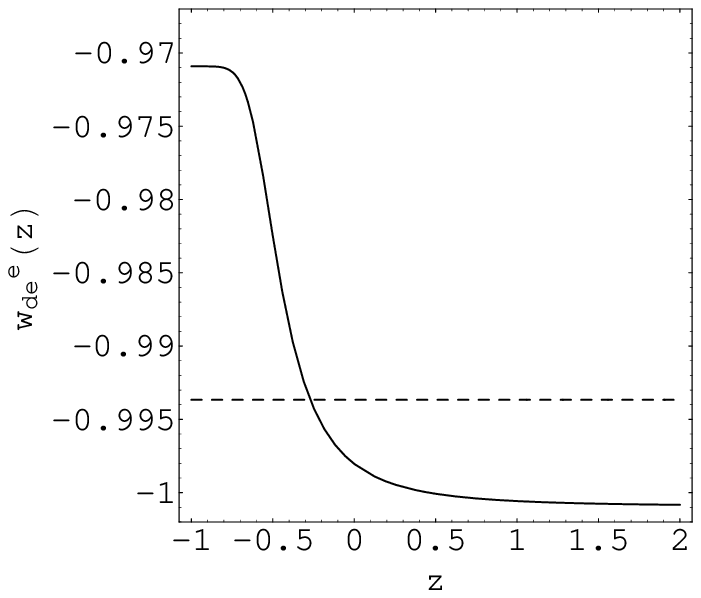}~~
   \includegraphics[width=5.5cm]{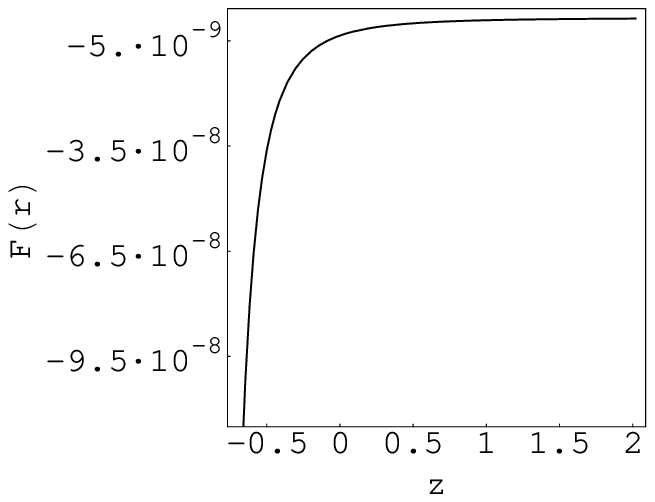}\\
  \caption{The evolutions of the effective state parameter for dark energy
  $w_{de}^{e}(z)$ (left)
   in interacting model with constant function
  $F(r)$ (dot line) and variable function  $F(r)$ (solid line), and the evolution of
  variable interactioin  function $F(r)$ (right).}\label{figure-weff-Fr-557+4}
\end{figure}

In addition, for seeing the influence from the interaction function
we also plot the evolutions of  the effective  state parameter
 for dark
 energy $w_{de}^{e}=\gamma_{de}^{e}-1$ and the interaction function $F(r)$. The
effective  state parameter
 for dark energy in above two interacting
scenarios are respectively expressed as
 \begin{eqnarray}
 w_{de}^{e}&&{=}\gamma_{de}^{e}-1\nonumber\\
 &&{=}\gamma_{de}+r\gamma_{dm}+\frac{\dot{\rho}_{dm}}{3H\rho_{de}}-1\nonumber\\
 &&{=}-1+\gamma_{de}+(\gamma_{dm}-\frac{\beta}{3})[r_{\infty}+(r_{0}-r_{\infty})
 (1+z)^{\frac{3(\gamma_{dm}-\gamma_{de})}{1+r_{\infty}}}]-\frac{\gamma_{dm}-\gamma_{de}}{1+r_{\infty}}
 (r_{0}-r_{\infty})
 (1+z)^{\frac{3(\gamma_{dm}-\gamma_{de})}{1+r_{\infty}}}\label{weff-dm-Frc}
 \end{eqnarray}
 for the interacting model with a constant function $F(r)$, and
 \begin{eqnarray}
 w_{de}^{e}=-1+\gamma_{de}-\frac{r_{\infty}^{2}\mu}{6}
 +(\gamma_{dm}-\nu)\sqrt{r_{\infty}^{2}+(r_{0}^{2}-r_{\infty}^{2})(1+z)^{\mu}}
 -\frac{\mu(r_{0}^{2}-r_{\infty}^{2})(1+z)^{\mu}}{6\sqrt{r_{\infty}^{2}
 +(r_{0}^{2}-r_{\infty}^{2})(1+z)^{\mu}}}\label{weff-dm-Frv}
 \end{eqnarray}
 for the interacting model with a variable function $F(r)$.
 By using the best fit values of model parameters from the combined  constraint of
 SNIa+OHD+CBF+BAO+CMB data, where $r_{\infty}=-0.006$ and $\gamma_{de}=0.012$,
  the evolutions of  effective
  state parameter for dark energy $w_{de}^{e}(z)$ in above two interacting
  scenarios are plotted in Fig. \ref{figure-weff-Fr-557+4} (left). From
  this figure one can see that for the case of constant interaction
  function, the $w_{de}^{e}(z)$ is almost constant; and for the case of
  variable interaction function, the parameter $w_{de}^{e}(z)$
  slowly change with respect to the redshift $z$. Furthermore  for the
  interaction function $F(r)$, we have the best fit value of
  $F(r)=-\frac{r_{\infty}}{1+r_{\infty}}(\gamma_{dm}-\gamma_{de})=0.006$ for the constant interacting case,
  and plot the best fit evolution of
 $F(r)=-\frac{(1-r)r_{\infty}^{2}}{r(1-r_{\infty}^{2})}(\gamma_{dm}-\gamma_{de})$
 in Fig. \ref{figure-weff-Fr-557+4} (right) for the variable interaction function. From
 Fig. \ref{figure-weff-Fr-557+4} (right)
 it is easy to see that the interaction between dark matter
 and dark energy is always very weak, though it is variational with respect to redshift $z$.

\section{$\text{The evolutions of geometrical quantities with their confidence level}$}

In this part we investigate the evolutions of  some geometrical
quantities with their  confidence level, such as deceleration
parameter $q(z)$ and jerk parameter $j(z)$.
 The confidence level on a function $f =
f(\theta)$ in terms of the variables $\theta$ are calculated by
\begin{equation}
 \sigma_{f}^{2}=\sum_{i}^{m}(\frac{\partial f}{\partial
  \theta_{i}})^{2}C_{ii}+2\sum_{i}^{m}\sum_{j=i+1}^{m}
  (\frac{\partial f}{\partial \theta_{i}})(\frac{\partial f}{\partial
  \theta_{j}})C_{ij},
\end{equation}
where $m$ is the number of parameters, $\theta$ denotes model
parameters, $C_{ij}$ is the  covariance matrix of the fitting
parameters that is the inverse of the Fisher matrix $(C_{ij}^{-1})
=\frac{1}{2}\frac{\partial^{2}\chi^{2}(\theta)}{\partial
\theta_{i}\partial \theta_{j}}$,
  $f(z; \theta_{i})$ express
any one cosmological parameter.  The evolution of any cosmological
quantity $f(z)$ with confidence level is given by
\begin{equation}
f_{1\sigma}(z)=f(z)\mid_{\theta=\bar{\theta}}\pm \sigma _{f},
\end{equation}
here $\bar{\theta}$ is the best fit values of the constraint
parameters.

The deceleration parameter is defined as
\begin{equation}
q(z)\equiv -\frac{\ddot{a}}{aH^{2}}=(1+z)\frac{1}{H}\frac{dH}{dz}-1.
\end{equation}
For the case of constant interaction function $F(r)$, one has
\begin{equation}
q(z)=-1+\frac{-3
r_{\infty}(1-\Omega_{0m})(1+z)^{3}+3\Omega_{0m}(1+z)^{3}+\beta(1+r_{\infty})(1-\Omega_{0m})(1+z)^{\beta}}
{2[-r_{\infty}(1-\Omega_{0m})(1+z)^{3}+\Omega_{0m}(1+z)^{3}+(1+r_{\infty})(1-\Omega_{0m})(1+z)^{\beta}]}.
\end{equation}
For $r_{\infty}=0$, it reduces to the non-interacting case. For the
case of variable interaction function $F(r)$,  the concrete form of
deceleration parameter is not listed here, since this expression is
too complex.
 In Fig. \ref{figure-qz-Fr} we plot the evolutions of
$q(z)$ for two interacting cases. According to the figures the
calculation results for transition redshift $z_{T}$ and current
deceleration parameter $q_{0}$ are listed in table \ref{table-q-Fr}.
From table \ref{table-q-Fr}, comparing two interacting scenarios it
can been seen that the constant interacting model tends to have the
smaller values of transition redshift $z_{T}$ and the more violent
decelerated-expansion rhythm at present (reflected by the smaller
value of $q_{0}$). And  from Fig. \ref{figure-qz-Fr}, it is easy to
see that for the case of interacting model with a variable function
$F(r)$, it has the more stringent constraints on the evolutions of
deceleration parameter than the case of interacting model with a
constant function $F(r)$.

\begin{figure}[ht]
   \includegraphics[width=4cm]{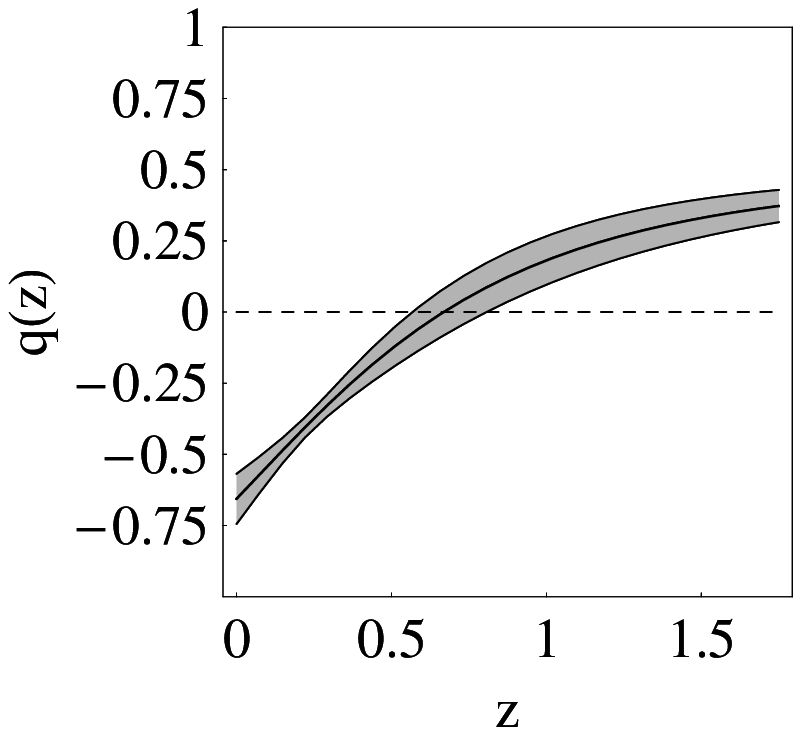}~~~~
    \includegraphics[width=4cm]{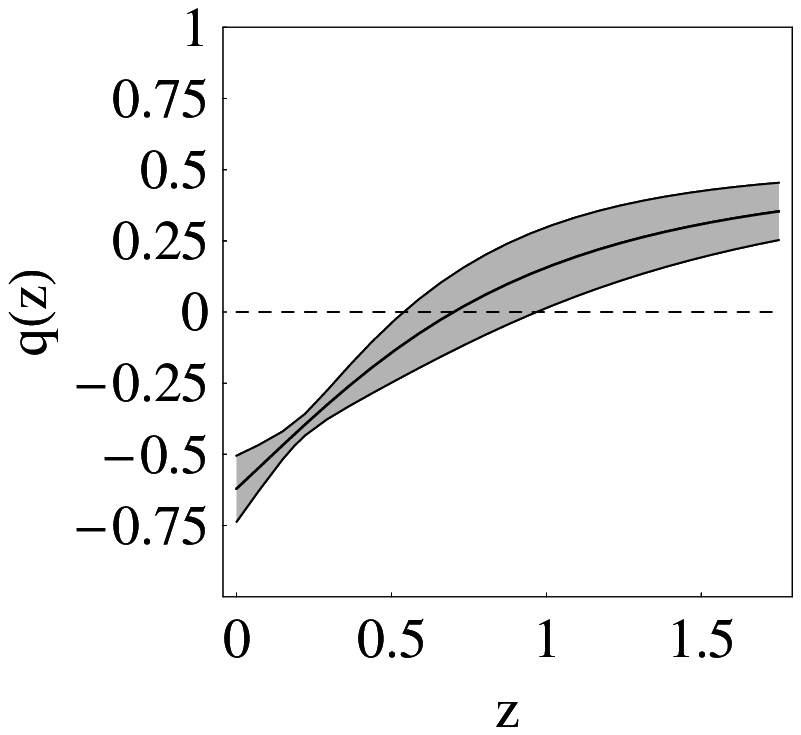}~~~~
    \includegraphics[width=4cm]{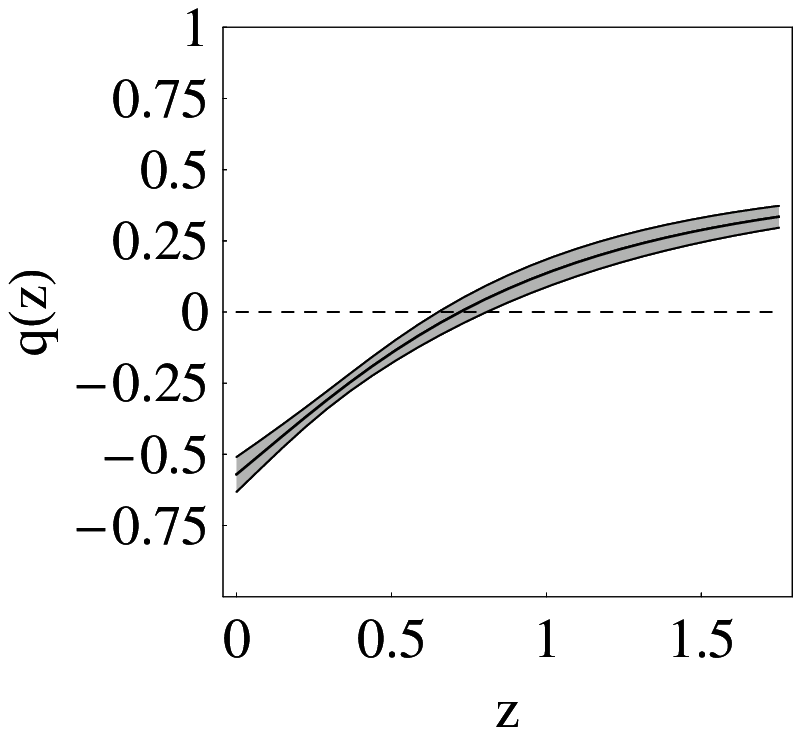}\\
    \includegraphics[width=4cm]{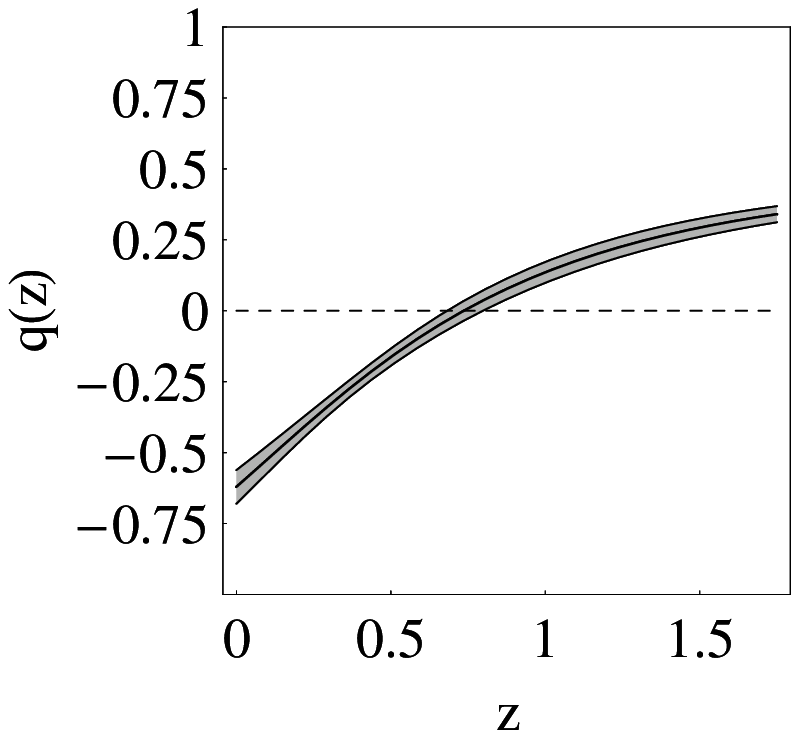}~~~~
   \includegraphics[width=4cm]{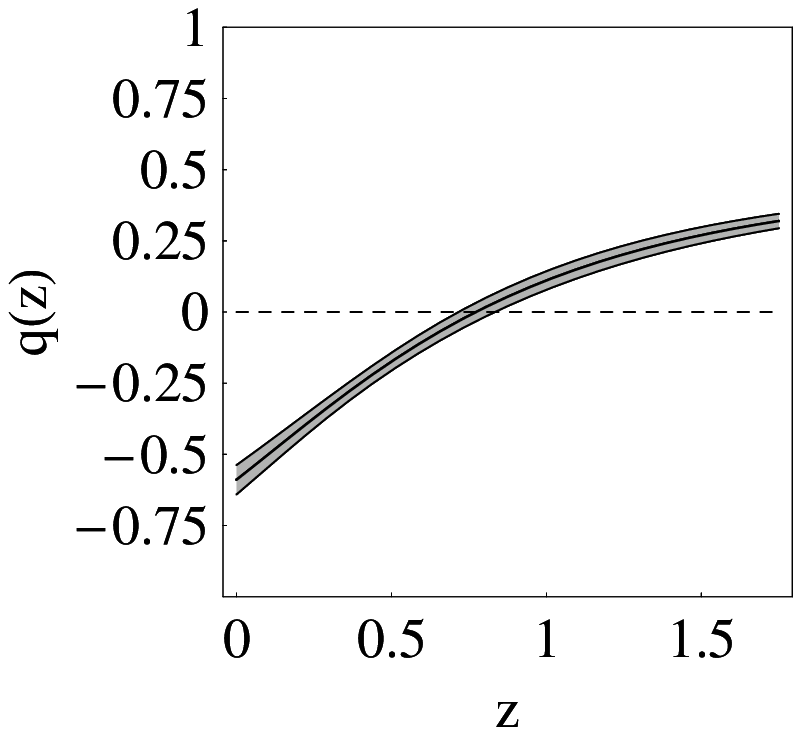}~~~~
   \includegraphics[width=4cm]{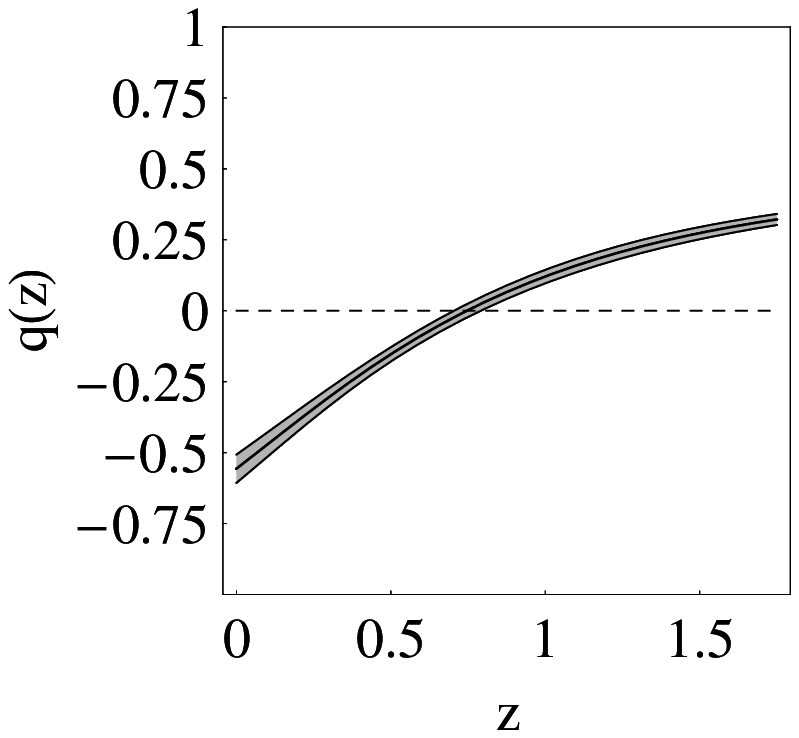}\\
  \caption{The evolutions of $q(z)$ for interacting model with constant function
  $F(r)$ (upper) and variable function $F(r)$ (down),
  from the combined observational data of  SNIa+OHD+BAO (left), SNIa+CMB (middle) and
   SNIa+OHD+CBF+BAO+CMB (right).}\label{figure-qz-Fr}
\end{figure}

\begin{table}[ht]
 \vspace*{-12pt}
 \begin{center}
 \begin{tabular}{c |  c|  c|| c |  c  } \hline\hline
                  & $z_{T}$ &  $q_{0}$ & $z_{T}$ &  $q_{0}$   \\\hline
 SNIa+OHD+BAO     & $0.669^{+0.137}_{-0.101}$   & $-0.657^{+0.088}_{-0.087}$
                  &  $0.737^{+0.061}_{-0.054}$  & $-0.621^{+0.059}_{-0.059}$ \\\hline
 SNIa+CMB         & $0.701^{+0.267}_{-0.162}$ & $-0.621^{+0.166}_{-0.166}$
                   & $0.773^{+0.060}_{-0.056}$   & $-0.589^{+0.051}_{-0.052}$ \\\hline
 SNIa+OHD+CBF+BAO+CMB & $0.721^{+0.085}_{-0.070}$  & $-0.571^{+0.061}_{-0.060}$
                   & $0.747^{+0.044}_{-0.042}$& $-0.557^{+0.050}_{-0.050}$\\\hline\hline
 \end{tabular}
 \end{center}
 \caption{  The values of transition redshift  $z_{T}$ and current deceleration parameter $q_{0}$
  for the interacting model with the constant function $F(r)$ (left) and the variable function $F(r)$
  (right). }\label{table-q-Fr}
 \end{table}


The jerk parameter is defined by scale factor $a$ and its third
derivative \cite{jerkvalue,jerkvalue1,jerkvalue2},
\begin{equation}
j \equiv
-\frac{1}{H^{3}}(\frac{\dot{\ddot{a}}}{a})=-[\frac{1}{2}(1+z)^{2}\frac{[H(z)^{2}]^{''}}{H(z)^{2}}
-(1+z)\frac{[H(z)^{2}]^{'}}{H(z)^{2}}+1].\label{jerk}
\end{equation}
 The use of the cosmic jerk parameter provides more
parameter space for geometrical  studies, and transitions between
phases of different cosmic acceleration are more naturally described
by models incorporating a cosmic jerk. Also, we list the expression
of jerk parameter for the case of constant interaction function
$F(r)$, with having a form
\begin{equation}
j =
-1-\frac{\beta(\beta-3)(1+r_{\infty})(1-\Omega_{0m})(1+z)^{\beta}}
{2[-r_{\infty}(1-\Omega_{0m})(1+z)^{3}+\Omega_{0m}(1+z)^{3}+(1+r_{\infty})(1-\Omega_{0m})(1+z)^{\beta}]}.\label{jerk-Fc}
\end{equation}
 For the
evolutions of jerk parameter $j(z)$ in interacting models including
the cases of constant function and variable function are plotted in
Fig. \ref{figurejz-557+4} by using the combined observational data
of SNIa+OHD+CBF+BAO+CMB. The current values of jerk parameter for
the cases of constant and variable interaction function  are
respectively given by, $j_{01}=-0.980^{+0.253}_{-0.252}$ and
$j_{02}=-0.906^{+0.156}_{-0.143}$. For the case of the
non-interacting model-independent scenario,  the evolutions of
deceleration parameter $q(z)$ and jerk parameter $j(z)$, and the
detailed discussions on the current values of deceleration parameter
$q_{0}$ and jerk parameter $j_{0}$ can be found in Ref.
\cite{cosmography-prd}, where the combined constraint results are
obtained from the latest observational data, according to the
analysis of Cosmography.

\begin{figure}[ht]
   \includegraphics[width=4cm]{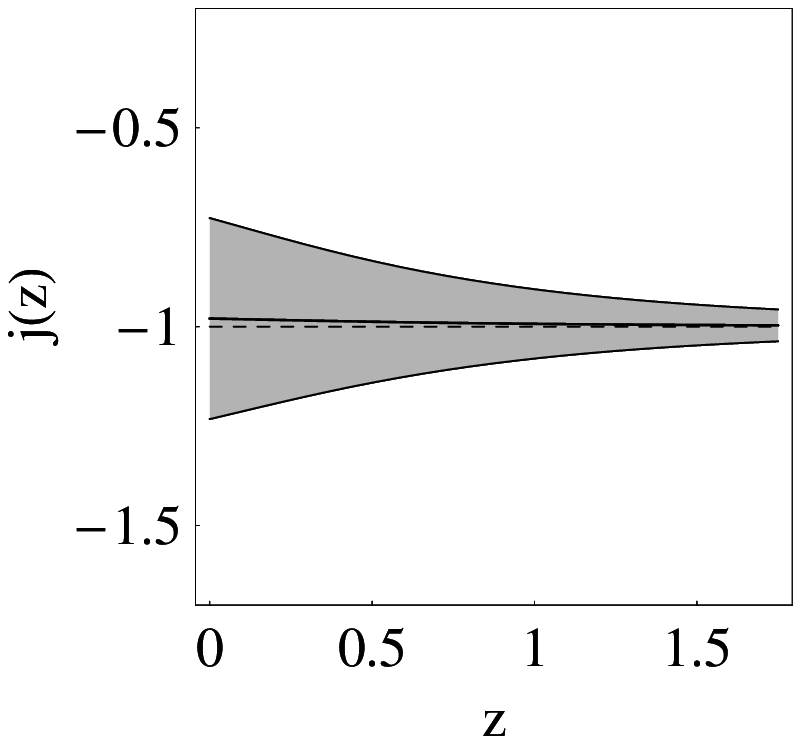}~~~~~~~~~
   \includegraphics[width=4cm]{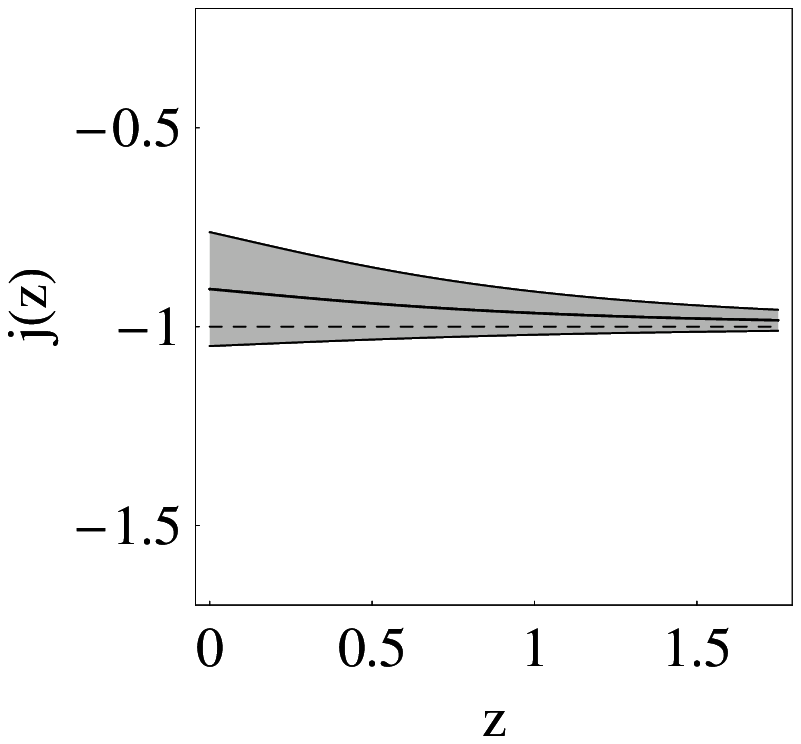}\\
  \caption{The evolution of jerk parameter $j(z)$  for interacting model with constant function
  $F(r)$ (left) and variable function  $F(r)$ (right)
   from the combined data: SNIa+OHD+CBF+BAO+CMB.}\label{figurejz-557+4}
\end{figure}


\section{$\text{Conclusions}$}

One knows the popular interpretation to the accelerating universe is
the cosmological constant model. But this model suffers from the
fine-tinning and the coincidence problems. And one of solutions to
solve these problems is to consider the interaction between two dark
sections. In this paper, following Ref. \cite{Fr}  we investigate
the interaction with using two different methods. We apply the
current observed data, including 557 Union2 SNIa, OHD, cluster X-ray
gas mass fraction, BAO and CMB data, to constrain the interacting
dark models with considering the constant interaction function and
the variable interaction function $F(r)$. According to the
constraint results on model parameters,  it indicates that the
interaction between dark matter and dark energy is occurred, but the
interacting strength is weak for the case of the constant function
$F(r)$. When consider the interaction with the variable function, it
seems that the interaction between dark matter and dark energy is
not obvious for the best fit analysis. In addition, we consider the
evolution of geometrical quantities, such as deceleration parameter
and jerk parameter. It is shown that the most stringent constraint
on deceleration parameter is given by
 the combined constraint of SNIa+OHD+CBF+BAO+CMB data. And we also
 get the constraint results on some cosmological quantities, such as transition
 redshift,  current deceleration parameter and jerk parameter.

 For the analysis of effective state parameter for dark energy $w_{de}^{e}$
 and  dark-matter energy density $\rho_{dm}$, we consider using the best fit model parameters from
 the combined constraint of SNIa+OHD+CBF+BAO+CMB data. From Fig. \ref{figure-weff-Fr-557+4} (left) it is
 shown that for the case of the variable interaction function, due to the influence of
 interaction between dark matter and dark energy  the parameter $w_{de}^{e}$ is dynamical,
 but the evolution is
  slow in the future ($z<0$), and
  go near to be constant in the past ($z>0$).
  For the case of the variable interaction function, $w_{de}^{e}$ is
 almost constant all the time.
 Furthermore according to the best fit values of parameters
 $\alpha=\frac{(\gamma_{dm}-\gamma_{de})}{1+r_{\infty}}\simeq 0.982$
 and
 $\beta=3\frac{r_{\infty}\gamma_{dm}+\gamma_{de}}{1+r_{\infty}}\simeq
 0.019$ for the  constant interacting model,
 $\mu=\frac{6(\gamma_{dm}-\gamma_{de})}{1-r_{\infty}^{2}}\simeq
 5.820$ and
$\nu=\gamma_{de}-\frac{r_{\infty}^{2}(\gamma_{dm}-\gamma_{de})}{1-r_{\infty}^{2}}\simeq
0.030$ for the  variable interacting model, from Eq. (\ref{rho1})
and (\ref{rhodm-Fv}) it can be found that for these two interacting
models the evolutions of dark matter obey, $\rho_{dm1}\propto
a^{-3\alpha-\beta}\simeq a^{-3.003}$ and $\rho_{dm2}\propto
a^{-\mu/2+3\nu}\simeq a^{-3}$, which is similar  to the popular
understanding $\rho_{dm}\propto a^{-3}$. Then the
acceleration-expanded universe will not appear in the
matter-dominated phase  for the interacting models (as shown in
Fig.\ref{figure-qz-Fr} about the evolution of deceleration parameter
$q$), which is a fundamental for the  structure formation. In
addition, we note that according to the second law of thermodynamics
\cite{second-law},
  it requires that the energy density is transferred from dark energy to dark
  matter.  From the analysis of interaction function, it
 is shown that  $F(r)$ should be smaller than zero.
  This condition is satisfied for the case of variable interaction function
  (one can see in Fig. \ref{figure-weff-Fr-557+4} (right)).
  For the case of constant interaction function, though the best fit
  value   of  $F(r)$ is not satisfied (since $r_{\infty}<0$ ), it comes into existence at $1\sigma$
  and $2\sigma$ confidence levels.

 \textbf{\ Acknowledgments}
 The research work is supported by   the National Natural Science Foundation of
China (11147150), the Natural Science Foundation of Education
Department of Liaoning Province (L2011189), the Natural Science
Foundation of Liaoning Province (Grant No.20102124),
 the NSFC (11175077) and the NSFC (11005088)  of P.R. China.

\end{document}